\documentclass[conference]{IEEEtran}
\usepackage{cite}
\usepackage{soul}
\usepackage{url}
\usepackage{amsmath}
\usepackage[utf8]{inputenc}
\usepackage[english]{babel}
\usepackage{amsthm}
\usepackage{amsfonts}
\usepackage{graphicx}  
\usepackage{algorithm}
\usepackage{algpseudocode}
\usepackage{caption}
\usepackage{subcaption}
\usepackage{multirow}

\DeclareMathOperator*{\argmax}{argmax}
\begin{document}
\title{Insider Threat Detection via Hierarchical Neural Temporal Point Processes}
\author{\IEEEauthorblockN{Shuhan Yuan, Panpan Zheng, Xintao Wu, Qinghua Li}
\IEEEauthorblockA{University of Arkansas, Fayetteville, AR, USA\\
Email: \{sy005, pzheng, xintaowu, qinghual\}@uark.edu}
}
\maketitle
\begin{abstract}
Insiders usually cause significant losses to organizations and are hard to detect. Currently, various approaches have been proposed to achieve insider threat detection based on analyzing the audit data that record information of the employee's activity type and time. However, the existing approaches usually focus on modeling the users' activity types but do not consider the activity time information. In this paper, we propose a hierarchical neural temporal point process model by combining the temporal point processes and recurrent neural networks for insider threat detection. Our model is capable of capturing a general nonlinear dependency over the history of all activities by the two-level structure that effectively models activity times, activity types, session durations, and session intervals information. Experimental results on two datasets demonstrate that our model outperforms the models that only consider information of the activity types or time alone.
\end{abstract}

\section{Introduction}
An insider threat is a malicious threat from people within the organization.  It may involve intentional fraud, the theft of confidential or commercially valuable information, or the sabotage of computer systems.
The subtle and dynamic nature of insider threats makes detection extremely difficult. The 2018 U.S. State of Cybercrime Survey indicates that 25\% of the cyberattacks are committed by insiders, and 30\% of respondents indicate incidents caused by insider attacks are more costly or damaging than outsider attacks \cite{Forcepoint20182018}.


Various insider threat detection approaches have been proposed \cite{eldardiry2013multi,Rashid2016New,Le2018Evaluating,Salem2008Survey,Sanzgiri2016Classification,Tuor2017Deep,Senator:2013}. However, most of the existing approaches only focus on operation type (web visit, send email, etc) information and do not consider the crucial activity time information.
In this paper, we study how to develop a detection model that captures both activity time and type information.
In literature, the marked temporal point process (MTPP) is a general mathematical framework to model the event time and type information of a sequence. It has been widely used for predicting the earthquakes and aftershocks \cite{Reinhart2017Review}. The traditional MTPP models make assumptions about how the events occur, which may be violated in reality. Recently, researchers \cite{Du2016Recurrent,Mei2017Neural} proposed to combine the temporal point process with recurrent neural networks (RNNs). Since the neural network models do not need to make assumptions about the data, the RNN-based MTPP models usually achieve better performance than the traditional MTPP models.

However, one challenge of applying RNN-based temporal point processes in insider threat detection is it cannot model the time information in multiple time scales. For example, user activities are often grouped into sessions that are separated by operations like ``LogOn'' and ``LogOff''. The dynamics of activities within sessions are different from the dynamics of sessions.  To this end, we propose a hierarchical RNN-based temporal point process model that is able to capture both the intra-session and inter-session time information.
Our model contains two layers of long short term memory networks (LSTM) \cite{Hochreiter1997Long}, which are variants of the traditional RNN.  The lower-level LSTM captures the activity time and types in the intra-session level, while the upper-level LSTM captures the time length information in the inter-session level. In particular, we adopt a sequence to sequence model in the lower-level LSTM, which is trained to predict the next session given the previous session. The upper-level LSTM takes the first and last hidden states from the encoder of the lower-level LSTM as inputs to predict the interval of two sessions and the duration of next session.  By training the proposed hierarchical model with the activity sequences generated by normal users, the model can predict the activity time and types in the next session by leveraging the lower-level sequence to sequence model, the  time interval between two consecutive sessions and the session duration time from the upper-level LSTM.  In general, we expect our model trained by normal users can predict the normal session with high accuracy. If there is a significant difference between the predicted session and the observed session, the observed session may contain malicious activities from insiders.

Our work makes the following contributions: (1) we develop an insider threat detection model that uses both activity type and time information; (2) we propose a hierarchical neural temporal point process model that can effectively capture two time-scale information; (3) the experiments on two datasets demonstrate that combining the activity type and multi-scale time information achieves the best performance for insider threat detection.


\section{Related Work}
\subsection{Insider Threat Detection}
Much of the research work on the characterization of insiders. Based on the intention of the attack, there are three types of insiders, i.e., \textit{traitors} who misuse his privileges to commit malicious activities, \textit{masqueraders} who conduct illegal actions on behalf of legitimate employees of an institute, and \textit{unintentional perpetrators} who unintentionally make mistakes \cite{Liu2018Detecting}. Based on the malicious activities conducted by the insiders, the insider threats can also be categorized into three types, \textit{IT sabotage} which indicate to directly uses IT to make harm to an institute, \textit{theft of intellectual property} which indicates to steal information from the institute, \textit{fraud} which indicates unauthorized modification, addition, or deletion of data \cite{Homoliak:2019}.

It is well accepted that the insiders' behaviors are different from the behaviors of legitimate employees. Hence, analyzing the employees' behaviors via the audit data plays an important role in detecting insiders. In general, there are three types of data sources, host-based, network-based, and context data. The host-based data record activities of employees on their own computers, such as command lines, mouse operations and etc. The network-based data indicate the logs recorded by network equipment such as routers, switches, firewalls and etc. The context data indicate the data from an employee directory or psychological data.

Given different types of data sources, various insider threat detection algorithms have been proposed. For example, some researchers propose to adopt decoy documents or honeypots to lure and identify the insiders \cite{1254322}. Meanwhile, one common scenario is to consider the insider threat detection as an anomaly detection task and adopt the widely-used anomaly detection approaches, e.g., one-class SVM, to detect the insider threats \cite{Sanzgiri2016Classification}. 
Moreover, some approaches treat the employee's actions over a period of time on a computer  as a sequence. The sequences that are frequently observed are normal behavior, while the sequences that are seldom observed are abnormal behavior that could be from insiders.  Research in \cite{Rashid2016New} adopts Hidden Markov Models (HMMs) to learn the behaviors of normal employees and then predict the probability of a given sequence. An employee activity sequence with low probability predicted by HMMS could indicate an abnormal sequence.  Research in \cite{Le2018Evaluating} evaluates an insider threat detection workflow using supervised and unsupervised learning algorithms, including Self Organizing Maps (SOM), Hidden Markov Models (HMM), and Decision Trees (DT). However, the existing approaches do not model the activity time information. In this work, we aim to capture both the activity time and type information for insider threat detection.

\subsection{Temporal Point Process}
A temporal point process (TPP) is a stochastic process composed of a time series of events that occur in continuous time \cite{Rasmussen2018Lecture}. The temporal point process is widely used for modeling the sequence data with time information, such as health-care analysis, earthquakes and aftershocks modeling and social network analysis \cite{Zhao2015Seismic,Reinhart2017Review,Farajtabar2018Point}. The traditional methods of temporal point processes usually make parametric assumptions about how the observed events are generated, e.g., by Poisson processes or self-exciting point processes. If the data do not follow the prior knowledge, the parametric point processes may have poor performance. To address this problem, researchers propose to learn a general representation of the dynamic data  based on neural networks without assuming parametric forms \cite{Du2016Recurrent,Mei2017Neural}. Those models are trained by maximizing log likelihood. Recently, there are also emerging works incorporating the objective function from generative adversarial network \cite{Xiao2017Wasserstein,Zha2018Improving} or reinforcement learning \cite{Li2018Learning} to further improve the model performance. However, the current TPP models only focus on one granularity of time. In our scenario, we propose a hierarchical RNN framework to model the multi-scale time information.

\section{Preliminary}
\subsection{Marked Temporal Point Process}
Marked temporal point process is to model the observed random event patterns along time. A typical temporal point process is represented as an event sequence $S =\{e_1, \cdots, e_j, \cdots, e_T\}$. Each event $e_j=(t_j, a_j)$ is associated with an activity type $a_j \in \mathcal{A}=\{1,\cdots,A\}$ and an occurred time $t_j \in [0,T]$. Let $f^*((t_j,a_j)) = f((t_j,a_j)|\mathcal{H}_{t_{j-1}})$ be the conditional density function of the event $a_j$ happening at time $t_j$ given the history events up to time $t_{j-1}$, where $\mathcal{H}_{t_{j-1}}=\{(t_{j'}, a_{j'})|t_{j'}<=t_{j-1}, a_{j'} \in \mathcal{A}\}$ as the collected historical events before time $t_j$. Throughout this paper, we use $*$ notation to denote that the function depends on the history. The joint likelihood of the observed sequence $S$ is:
\begin{equation}
    f \big(\{(t_j,a_j)\}_{j=1}^{|S|} \big)=\prod_{j=1}^{|S|} f((t_j,a_j)|\mathcal{H}_{t_{j-1}}) = \prod_{j=1}^{|S|} f^*((t_j,a_j)).
\end{equation}
There are different forms of $f^*((t_j,a_j))$. However, for mathematical simplicity, it usually assumes the times $t_j$ and mark $a_j$ are conditionally independent given the history $\mathcal{H}_{t_{j-1}}$, i.e., $f^*((t_j,a_j)) = f^*(t_j)f^*(a_j)$, where $f^*(a_j)$ models the distribution of event types; $f^*(t_j)$ is the conditional density of the event occurring at time $t_j$ given the timing sequences of past events \cite{Du2016Recurrent}.

A temporal point process can be characterized by the \textit{conditional intensity function}, which indicates the expected instantaneous rate of future events at time $t$:
\begin{equation}
\label{eq:lambda}
    \lambda^*(t)=\lambda(t|\mathcal{H}_{t_{j-1}})=\lim_{\text{d}t \rightarrow 0} \frac{\mathbb{E}[N([t,t+\text{d} t])|\mathcal{H}_{t_{j-1}}]}{\text{d}t},
\end{equation}
where $N([t,t+\text{d} t])$ indicates the number of events occurred in a time interval $\text{d} t$. Given the conditional density function $f$ and the corresponding cumulative distribution $F$ at time $t$, the intensity function can be also defined as:
\begin{equation}
\label{eq:lambda_f}
    \lambda^*(t)=\frac{f(t|\mathcal{H}_{t_{j-1}})}{S(t|\mathcal{H}_{t_{j-1}})}=\frac{f(t|\mathcal{H}_{t_{j-1}})}{1-F(t|\mathcal{H}_{t_{j-1}})},
\end{equation}
where $S(t|\mathcal{H}_{t_{j-1}})=exp(-\int_{t_{j-1}}^t \lambda^*(\tau)d\tau)$ is the \textit{survival function} that indicates the probability that no new event has ever happened up to time $t$ since $t_j-1$. Then, the \textit{conditional density function} can be described as:
\begin{equation}
\label{eq:f}
    f^*(t)=f(t|\mathcal{H}_{t_{j-1}})=\lambda^*(t)exp\Big(-\int_{t_{j-1}}^t \lambda^*(\tau)d\tau \Big).
\end{equation}

Particular functional forms of the conditional intensity function $\lambda^*(t)$ for Poisson process, Hawkes process, self-correcting process, and autoregressive conditional duration process have been widely studied \cite{Rasmussen2018Lecture}.
For example, a Hawkes process captures the self-excitation phenomenon among events \cite{hawkes1971spectra}. The conditional intensity function of a Hawkes process is defined as:
\begin{equation}
    \lambda^*(t) = \lambda_0 + \sum_{t_j \in \mathcal{H}_{t}} \gamma(t,t_j),
\end{equation}
where $\lambda_0>0$ is the base intensity that indicates the intensity of events triggered by external signals instead of previous events; $\gamma(t,t_j)$ is the triggering kernel which is usually predefined as $\gamma(t,t_j) = \exp(\beta (t-t_j))$. The Hawkes process models the self-excitation phenomenon that the arrival of an event increases the conditional intensity of observing events in the near future. Recently, the Hawkes process is widely used to model the information diffusion on online social networks. 

However, these different parameterization models make different assumptions about the latent dynamics that is never known in practice. For example, the self-excitation assumption of the Hawkes process may not be held in many scenarios. The model misspecification can seriously degrade the predictive performance.

\subsection{Sequence-to-Sequence Model}
In general, a sequence-to-sequence (seq2seq) model is used to convert sequences from one domain to sequences in another domain. The seq2seq consists of two components, one encoder and one decoder. Both encoder and decoder are long short-term memory (LSTM) models and can model the long-term dependency of sequences. The seq2seq model is able to encode a variable-length input to a fixed-length vector and further decode the vector back to a variable-length output. The length of the output sequence could be different from that of the input sequence.
The goal of the seq2seq model is to estimate the conditional probability $P(y_1, \cdots, y_{T'}|x_1, \cdots, x_T)$, where $(x_1, \cdots, x_T)$ is an input sequence and $(y_1, \cdots, y_{T'})$ is the corresponding output sequence. The \textbf{encoder} encodes the input sequence to a hidden representation with an LSTM model $\mathbf{h}^{en}_j = LSTM^{en} (\mathbf{x}_j, \mathbf{h}^{en}_{j-1})$ where $\mathbf{x}_j$ is the up-to-date input,  $\mathbf{h}^{en}_{j-1}$ is the previous hidden state, and $\mathbf{h}^{en}_j$ is the learned current hidden state.
The last hidden state $\mathbf{h}^{en}_T$ captures the information of the whole input sequence. The \textbf{decoder} computes the conditional probability $P(y_1, \cdots, y_{T'}|x_1, \cdots, x_T)$ by another LSTM model whose initial hidden state is set as $\mathbf{h}^{en}_T$:
\begin{equation}
 P(y_1, \cdots, y_{T'}|x_1, \cdots, x_T) = \prod_{j=1}^{T'} P(y_j|\mathbf{h}^{en}_T, y_1, \cdots, y_{j-1}).
\end{equation}
In seq2seq model, $P(y_j|\mathbf{h}^{en}_T, y_1, \cdots, y_{j-1}) = g(\mathbf{h}^{de}_j)$, where $\mathbf{h}^{de}_j=LSTM^{de} (\mathbf{y}_{j-1}, \mathbf{h}^{de}_{j-1})$ is the $j$-th hidden vector of the decoder; $g(\cdot)$ is usually a softmax function.


\begin{figure*}
    \centering
    \includegraphics[width=0.9\textwidth]{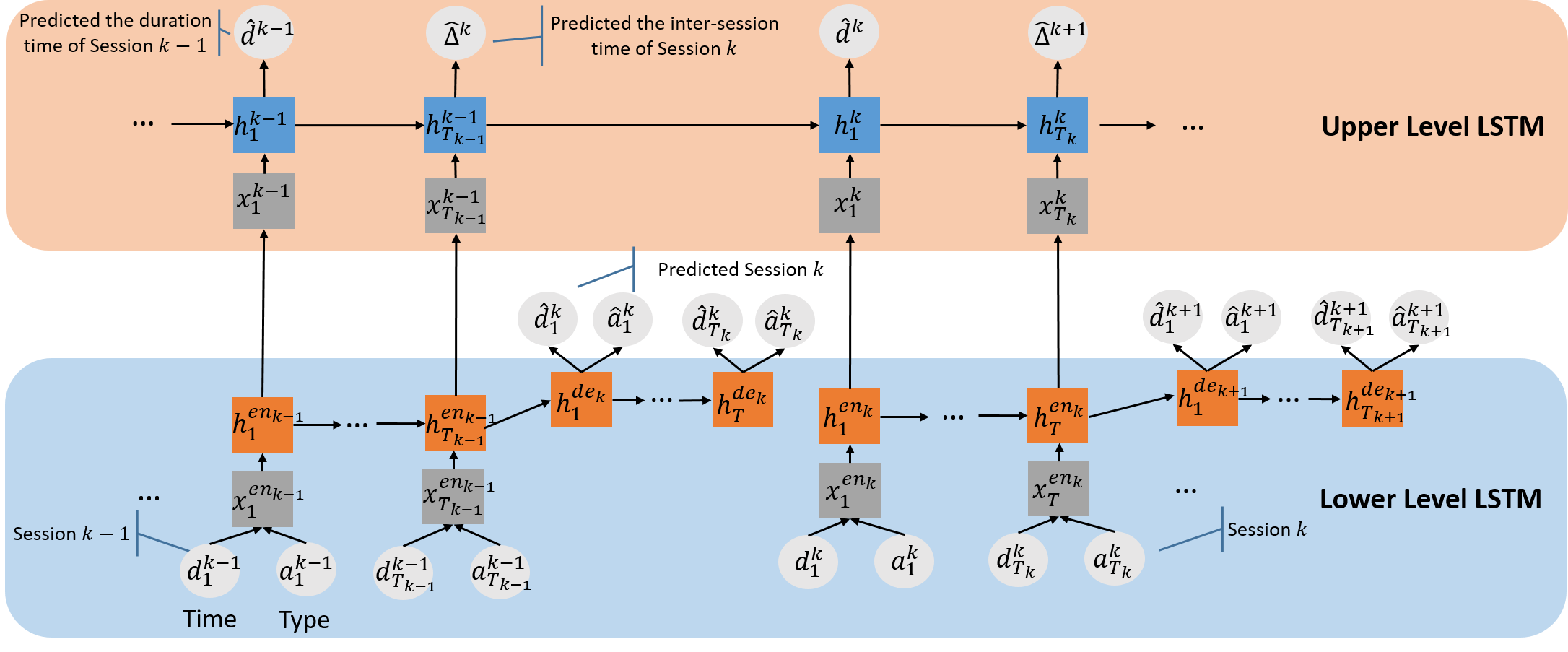}
    \caption{The framework for sequence generation with two time scales. The lower-level LSTM captures event patterns with the time and mark pairs in a session. The upper-level LSTM aims to predict the duration of sessions and inter-sessions.}
    \label{fig:framework}
\end{figure*}

\section{Insider Threat Detection}

\subsection{Framework}


We model a user's behavior as a sequence of activities that can be extracted from various types of raw data, such as user logins, emails, Web browsing, and FTP.  Formally, we model the up-to-date activities of a user as sequence $U=\{S^{1}, \cdots\, S^{k},\cdots\}$ where $S^{k}=\{e^k_1, \cdots, e^k_j, \cdots, e^k_{T_k}\}$ indicates his $k$-th activity session. For example, each session in our scenario is a sequence of activities starting with ``LogOn'' and ending with ``LogOff''.  $e^k_j=(t^{k}_j, a^{k}_j)$ denotes the $j$-th activity in the user's $k$-th session and contains activity type $a^{k}_j$ and occurred time $t^{k}_j$. We define $d^k_j = t^k_j-t^k_{j-1}$ as the inter-activity duration between activities $a^k_j$ and $a^k_{j-1}$,  $d^k=t^k_{T_k}-t^k_1$ as the length time of the $k$-th session, and $\Delta^k=t^{k}_1-t^{k-1}_{T_{k-1}}$ as the time interval between the $(k-1)$-th and $k$-th sessions. Note that $t^{k-1}_{T_{k-1}}$ is the occurred time of the last activity in the $(k-1)$-th session.

The goal of learning in our threat detection is to predict whether a new session $S^{k}=\{e^k_1, \cdots, e^k_j, \cdots, e^k_{T_k}\}$  is normal or fraudulent. To address the challenge that there are often no or very few records of known insider attacks for training our model, we propose a generative model that models normal user behaviors from a training dataset consisting of only sequences of normal users. The learned model is then used to calculate the fraudulent score of the new session $S^{k}$. We quantify the fraudulence of $S^{k}$ from two perspectives, activity information (including both type and time) within  sessions, and session time information (i.e., when a session starts and ends). For example, a user who foresees his potential layoff may have activities of uploading documents to Dropbox and visiting job-searching websites although he may try to hide these abnormal activities in multiple sessions; he may have ``LogOn'' and ``LogOff'' times different from his normal sessions as he may become less punctual or may have more sessions during weekends or nights, resulting different session durations and intervals between sessions. Moreover, when a user's account is compromised, activity and session information from the attacker will also be different even if the attacker tries to mimic the normal user's behaviors.


We develop a unified hierarchical model capable of capturing a general nonlinear dependency over the history of all activities.  Our detection model does not rely on any predefined signatures and instead use deep learning models to capture user behaviors reflected in raw data. Specifically, our hierarchical model learns the user behaviors in two time scales, intra-session level and inter-session level. For the intra-session level,
we adopt the seq2seq model
to predict $\hat{S}^{k}$ based on the previous $S^{k-1}$ and use the marked temporal point process model to capture the dynamic difference of activities.  Note that the number of activities of the predicted session $\hat{S}^{k}$ could be different from that of the previous $S^{k-1}$ as well as the true $S^k$. For the inter-session level, we aim to model the session interval $\Delta^k=t^{k}_1-t^{k-1}_{T_{k-1}}$  and the session duration $d^k=t^k_{T_k}-t^k_1$ of the $k$-th session.

The whole framework of predicting future events with two time scales is shown in Figure \ref{fig:framework}.
We do not assume any specific parametric form of the conditional intensity function. Instead, we follow \cite{Du2016Recurrent} to seek to learn a general representation to approximate the unknown dependency structure over the history. We also emphasize that the neural temporal point processes of two levels are connected in our framework.  The upper-level LSTM takes the first and last hidden states from the encoder of the lower-level LSTM as inputs to predict the interval of two sessions and the session duration. This connection guarantees the upper-level LSTM incorporates activity type information in its modeling.
For insider threat detection, since our model is trained by benign sessions, the predicted session $\hat{S}^{k}$ would be close to the observed $S^{k}$ when $S^{k}$ is normal, and different from $S^{k}$ when $S^{k}$ is abnormal. In Session \ref{session:metric}, we will present details about how to derive fraudulent score by comparing $(\hat{S}^{k},\hat{d}^{k},\hat{\Delta}^{k})$ with $(S^k, d^{k}, \Delta^{k})$, where $\hat{\bullet}$ indicates the predicted value.

\subsection{Intra-Session Insider Threat Detection }

In this work, we propose to use the seq2seq model to estimate the joint likelihood of $k$-th session  given the $(k-1)$-th session.
In particular, the encoder of the seq2seq model is to encode the activity time and type information at $(k-1)$-th session to a hidden representation. The decoder is to model the activity time interval $d^k_{j+1}$ and type $a^k_{j+1}$ information at $k$-th session given the history.

{\bf Encoder:} To map the $(k-1)$-th session to a hidden representation, the encoder first maps each activity occurring at time $t^{k-1}_j$ with type $a^{k-1}_j$ to an embedding vector $\mathbf{x}^{en_{k-1}}_j$:
\begin{equation}
\label{eq:input}
    \mathbf{x}^{en_{k-1}}_j = \mathbf{w}^t d^{k-1}_j + \mathbf{W}^{em} \mathbf{a}^{k-1}_j,
\end{equation}
where $d^{k-1}_j$ is the inter-activity duration between $a^{k-1}_{j}$ and $a^{k-1}_{j-1}$; $\mathbf{w}^t$ is a time-mapping parameter; $\mathbf{W}^{em}$ is an activity embedding matrix; $\mathbf{a}^{k-1}_j$ is a one-hot vector of the activity type $a^{k-1}_j$.
Then, by taking the entire sequence of $(k-1)$-th session as inputs to the encoder LSTM, the encoder projects the $(k-1)$-th session to a hidden representation $\mathbf{h}^{en_{k-1}}_{T_{k-1}}$.

{\bf Decoder:} The decoder is trained to predict the pairs of activity type and time at the $k$-th session given the information of $(k-1)$-th session. To predict the activity type information, given the hidden state of the decoder $\mathbf{h}^{de_{k}}_j$, the probability of the next activity having type value $a$ can be derived by a softmax function:
\begin{equation}
\label{eq:rnn_mark}
    P(a^k_{j+1}=a|\mathbf{h}^{de_k}_j) = \frac{exp(\mathbf{w}^s_a \mathbf{h}^{de_k}_j)}{\sum_{a'=1}^A exp(\mathbf{w}^s_{a'} \mathbf{h}^{de_k}_j)},
\end{equation}
where $\mathbf{w}^s_a$ is the $a$-th row of the weight matrix $\mathbf{W}^s$ in the softmax function.

To predict the activity time information, we adopt the conditional density function defined in Equation \ref{eq:f}. First, inspired by \cite{Du2016Recurrent}, we derive the LSTM-based conditional intensity function $\lambda^*(t)$ as:
\begin{equation}
\label{eq:rnn_lambda}
\lambda^*(t) = exp(\mathbf{v} \mathbf{h}^{de_k}_j + u^t(t-t^k_j)+b),
\end{equation}
where the exponential function is deployed to ensure the intensity function is always positive; $\mathbf{v}$ is a weight vector; $u^t$ and $b$ are scalars. Then, we can derive the conditional density function given the history until time $t^k_j$:
\begin{equation}
\begin{split}
\label{eq:rnn_f}
    &~ f^*(t) = \lambda^*(t) (\int_{t^k_j}^{t} \lambda^*(\tau)d\tau) \\
    &= exp\big(\mathbf{v}^t \mathbf{h}^{de_k}_j + u^t(t-t^k_j)+b^t + \frac{1}{u}exp(\mathbf{v}^t \mathbf{h}^{de_k}_j\\
    &+b^t)-\frac{1}{u}exp(\mathbf{v}^t \mathbf{h}^{de_k}_j + u^t(t-t^k_j)+b^t)\big).
\end{split}
\end{equation}

Hence, given the observed activity time information, we can calculate the conditional density function of the  time interval between two consecutive activities $d^k_{j+1}=t^k_{j+1} - t^k_j$ at $k$-th session:
\begin{equation}
\label{eq:f_t}
f^*(d^k_{j+1}) = f(d^k_{j+1}|\mathbf{h}^{de_k}_{j}).
\end{equation}

Since the lower-level LSTM is to model the time interval $d^k_{j+1}$ and type $a^k_{j+1}$ information, given a collection of activity sessions from benign employees, we combine the likelihood functions of the event type (Equation \ref{eq:rnn_mark}) and time (Equation \ref{eq:f_t}) to have the negative joint log-likelihood of the observation sessions:
\begin{equation}
\label{eq:session_likelihood}
\mathcal{L}_a = -\sum_{k=1}^M \sum_{j=1}^{T_k} \Big(\log P(a^k_{j+1}|\mathbf{h}_j^{de_k}) + \log f^*(d^k_{j+1}) \Big),
\end{equation}
where $M$ is the total number of sessions in the training dataset; $T_k$ is the number of activities in a session. The lower-level LSTM along with the decoder LSTM is trained by minimizing the negative log-likelihood shown in Equation \ref{eq:session_likelihood}.


When the model is deployed for detection, to obtain the predicted activity type $\hat{a}^k_{j+1}$, we simply choose the type with the largest probability $P(a|\mathbf{h}^{de_k}_j)$ (calculated by Equation \ref{eq:rnn_mark}):
\begin{equation}
    \hat{a}^k_{j+1} = \argmax_{a \in \mathcal{A}} P(a|\mathbf{h}^{de_k}_j).
\end{equation}
We further calculate the expected inter-activity duration between $(j+1)$-th and $j$-th activities $\hat{d}^k_{j+1} = E(t^k_j)$:
\begin{equation}
\label{eq:time_expectation}
\begin{split}
\hat{d}^k_{j+1} = \int_{t^k_{j}}^{\infty} t f^*(t) dt.
\end{split}
\end{equation}
The difference between $\hat{d}^k_{j+1}$ and the observed $d^k_{j+1}$ will be used to calculate the fraudulent score in terms of the timing information of intra-session activities.

\subsection{Inter-Session Insider Threat Detection}

The inter-session duration is crucial for insider threat detection.
To capture such information, we further incorporate an upper-level LSTM into the framework, which focuses on modeling the inter-session behaviors of employees. Specifically, the upper-level LSTM is trained to predict the inter-session duration between $k$-th and $(k-1)$-th sessions ($\Delta^k = t^k_1-t^{k-1}_{T_{k-1}}$) and the $k$-th session duration ($d^k = t^k_{T_k}-t^k_1$).

To predict the inter-session duration $\Delta^k$, the input of the upper-level LSTM  is from the last hidden state $\mathbf{h}^{en_{k-1}}_{T_{k-1}}$ of $(k-1)$-th session from the lower-level LSTM as shown in Equation \ref{eq:upper_input_1}, while to predict the $k$-th session duration $d^k$, the input of the upper-level LSTM is from the first hidden state $\mathbf{h}^{en_k}_1$ of $k$-th session as shown in Equation \ref{eq:upper_input_T}.
\begin{align}
    \label{eq:upper_input_1}
     \mathbf{x}^{{k-1}}_{T_{k-1}} &= \mathbf{U} \mathbf{h}^{en_{k-1}}_{T_{k-1}}, \\
     \label{eq:upper_input_T}
     \mathbf{x}^{k}_{1} &= \mathbf{U} \mathbf{h}^{en_k}_1,
\end{align}
where $\mathbf{U}$ is an input weight matrix for the upper-level LSTM.

Then, we can get the hidden states ($\mathbf{h}^{k-1}_{T_{k-1}}$ and $\mathbf{h}^{k}_1$) of the upper-level sequence based on an LSTM model.
Finally, the conditional density functions of the inter-session duration $\Delta^k$ and session duration $d^k$ are:
\begin{align}
f^*_s(\Delta^k) &= f(\Delta^k|\mathbf{h}^{k-1}_{T_{k-1}}), \\
f^*_s(d^{k}) &= f(d^{k}|\mathbf{h}^{k}_1),
\end{align}
where $f^*_s(\Delta^k)$ and $f^*_s(d^{k})$ can be calculated based on Equation \ref{eq:rnn_f}.

To train the upper-level LSTM, the negative log-likelihood of inter-session sequences can be defined as:
\begin{equation}
\label{eq:week_likelihood}
  \mathcal{L}_s = -\sum_{m=1}^{M'} \sum_{k=1}^{K_m} \big(\log f^*_s(\Delta^k_m) + \log f^*_s(d^k_m) \big),
\end{equation}
where $M'$ is the total number of inter-session level sequences in the training dataset; $K$ indicates the number of sessions in an inter-session level sequence. In our experiments, we use the upper-level LSTM to model the employee sessions in a week. Then, $M'$ indicates the total number of weeks in the training dataset, and $K$ is the number of sessions in a week. The upper-level LSTM is trained by minimizing the negative log-likelihood shown in Equation \ref{eq:week_likelihood}. After training, the upper-level LSTM can capture the patterns of the inter-session duration and session duration.

When the model is deployed for detection,  we calculate the predicted inter-session duration between $k$-th and ($k-1$)-th sessions $\hat{\Delta}^{k}$ and the $k$-th session duration $\hat{d}^k$, shown in Equations \ref{eq:d_delta}.
\begin{equation}
\label{eq:d_delta}
\hat{\Delta}^k = \int_{t^{k-1}_{T}}^{\infty} t f^*_s(t), \quad \hat{d}^k = \int_{t^k_{1}}^{\infty} t f^*_s(t) dt.
\end{equation}

The difference between $\hat{\Delta}^{k}$ ($\hat{d}^k$) and the observed $\Delta^{k}$ ($d^k$) will be used to calculate the fraudulent score in terms of the session timing information.

\subsection{Fraudulent Score}
\label{session:metric}

After obtaining the predicted session, we compare the generated times and types in a session with the observed session, respectively. For the activity types, we adopt the Bilingual Evaluation Understudy (BLEU) \cite{Papineni2002Bleu} score to evaluate the difference between the observed session and generated session. The BLEU metric was originally used for evaluating the similarity between a generated text and a reference text, with values closer to 1 representing more similar texts. BLEU is derived by counting matching n-grams in the generated text to n-grams in the reference text and insensitive to the word order. Hence, BLEU is suitable for evaluating the generated sequences and the observed sequences. We define the fraudulent score in terms of intra-session activity type as:
\begin{equation}
\label{eq:bleu}
    score_a = 1-BLEU(S^k_a, \hat{S}^k_a),
\end{equation}
where $S^k_a$ indicates the observed activity types in $k$-th session while $\hat{S}^k_a$ indicates the predicted session.
If $score_a$ is high, it means the observed session is a potentially malicious session in terms of session activity types.

For the activity time, as shown in Equation \ref{eq:mean_time}, we define the fraudulent score in terms of intra-session activity time by computing the mean absolute error (MAE) of the predicted time of each activity with the observed occurring time:
\begin{equation}
\label{eq:mean_time}
    score_t = \frac{1}{|S|} \sum^{|S|}_{j=1} |d^k_j-\hat{d}^k_j|.
\end{equation}

Since the upper-level LSTM takes each session's first and last hidden states as inputs to predict the time lengths of sessions and inter-sessions, we can further derive the time scores by comparing the predicted time lengths with the observed ones.
We define the fraudulent score in terms of inter-session duration as:
\begin{equation}
\label{eq:interval_time}
    score_{\Delta} = |\hat{\Delta}^{k}-\Delta^{k}|.
\end{equation}
Similarly, we define the fraudulent score in terms of session duration as:
\begin{equation}
\label{eq:session_time}
score_d = |\hat{d}^k-d^k|.
\end{equation}
Note that although $\hat{d}^k$ can be derived based on all the predicted activity time from lower-level LSTM, the error usually is high due to the accumulated error over the whole sequence. Hence, we use the upper-level LSTM to get the session time length. Finally, by combining Equations \ref{eq:bleu}, \ref{eq:mean_time}, \ref{eq:interval_time} and \ref{eq:session_time}, we define the total fraudulent score ($FS$) of a session as:
\begin{equation}
\label{eq:score}
    FS =  \alpha_1 score_a + \alpha_2 score_t + \alpha_3 score_d + \alpha_4 score_{\Delta},
\end{equation}
where $\alpha_1,\alpha_2,\alpha_3,\alpha_4$ are hyper-parameters, which can be set based on the performance of insider threat detection via using each score alone.

\section{Experiments}
\subsection{Experiment Setup}
{\bf \noindent Dataset.}
We adopt the CERT Insider Threat Dataset \cite{6565236}, which is the only comprehensive dataset publicly available for evaluating the insider threat detection. This dataset consists of five log files that record the computer-based activities for all employees, including \textbf{logon.csv} that records the logon and logoff operations of all employees, \textbf{email.csv} that records all the email operations (send or receive), \textbf{http.csv} that records all the web browsing (visit, download, or upload) operations, \textbf{file.csv} that records activities (open, write, copy or delete) involving a removable media device, and \textbf{decive.csv} that records the usage of a thumb drive (connect or disconnect). Table \ref{tb:types} shows the major activity types recorded in each file.
The CERT dataset also has the ground truth that indicates the malicious activities committed by insiders. We use the latest version (r6.2) of CERT dataset that contains 3995 benign employees and 5 insiders.

We join all the log files, separate them by each employee, and then sort the activities of each employee based on the recorded timestamps. We randomly select 2000 benign employees as the training dataset and another 500 employees as the testing dataset. The test dataset includes all sessions from  five insiders. The statistics of the training and testing datasets is shown in Table \ref{tb:data}. Based on the activities recorded in the log files, we extract 19 activity types shown in Table \ref{tb:types}. The activity types are designed to indicate the malicious activities.

\begin{table}[]
\centering
\caption{Statistics of Training and Testing Datasets}
\label{tb:data}
\resizebox{.45\textwidth}{!}{%
\begin{tabular}{|l|c|c|}
\hline
                  & Training Dataset & Testing Dataset \\ \hline
\# of Employees   & 2000             & 500              \\ \hline
\# of Sessions & 1039805          & 142600                \\ \hline \hline
\# of Insiders & 0               & 5               \\ \hline
\# of Malicious Sessions & 0              & 68              \\ \hline
\end{tabular}
}
\end{table}

\begin{table}[]
\centering
\caption{Operation Types Recorded in Log Files}
\label{tb:types}
\resizebox{.49\textwidth}{!}{
\begin{tabular}{|c|l|}
\hline
Files                       & \multicolumn{1}{c|}{Operation Types}                                                   \\ \hline
\multirow{4}{*}{logon.csv}  & Weekday Logon (employee logs on a computer on a weekday at work hours)                 \\ \cline{2-2}
                            & Afterhour Weekday Logon (employee logs on a computer on a weekday after work hours)    \\ \cline{2-2}
                            & Weekend Logon (employees logs on at weekends)                                          \\ \cline{2-2}
                            & Logoff (employee logs off a computer)                                                  \\ \hline
\multirow{4}{*}{email.csv}  & Send Internal Email (employee sends an internal email)                                 \\ \cline{2-2}
                            & Send External Email (employee sends an external email)                                 \\ \cline{2-2}
                            & View Internal Email (employee views an internal email)                                 \\ \cline{2-2}
                            & View external Email (employee views an external email)                                 \\ \hline
\multirow{3}{*}{http.csv}   & WWW Visit (employee visits a website)                                                  \\ \cline{2-2}
                            & WWW Download (employee downloads files from a website)                                 \\ \cline{2-2}
                            & WWW Upload (employee uploads files to a website)                                       \\ \hline
\multirow{4}{*}{device.csv} & Weekday Device Connect (employee connects a device on a weekday at work hours)         \\ \cline{2-2}
                            & Afterhour Weekday Device Connect (employee connects a device on a weekday after hours) \\ \cline{2-2}
                            & Weekend Device Connect (employee connects a device at weekends)                        \\ \cline{2-2}
                            & Disconnect Device (employee disconnects a device)                                      \\ \hline
\multirow{4}{*}{file.csv}   & Open doc/jpg/txt/zip File (employee opens a doc/jpg/txt/zip file)                      \\ \cline{2-2}
                            & Copy doc/jpg/txt/zip File (employee copies a doc/jpg/txt/zip file)                     \\ \cline{2-2}
                            & Write doc/jpg/txt/zip File (employee writes a doc/jpg/txt/zip file)                    \\ \cline{2-2}
                            & Delete doc/jpg/txt/zip File (employee deletes a doc/jpg/txt/zip file)                  \\ \hline
\end{tabular}
}
\end{table}

{\bf \noindent Baselines.}
We compare our model with two one-class classifiers: 1) One-class SVM \textbf{(OCSVM)} \cite{Tax2004Support} adopts support vector machine to learn a decision hypersphere around the positive data, and considers samples located outside this hypersphere as anomalies; 2) Isolation Forest (\textbf{iForest}) \cite{Liu2008Isolation} detects the anomalies with short average path lengths on a set of trees. For both baselines, we consider each activity type as an input feature and the feature value is the number of activities of the corresponding type in a session. In this paper, we do not compare with other RNN based insider threat detection methods (e.g., \cite{Tuor2017Deep}) as these methods were designed to detect the insiders or predict the days that contain insider threat activities.


{\bf \noindent Hyperparameters.} We map the extracted activity types to the type embeddings. The dimension of the type embeddings is 50. The dimension of the LSTM models is 100. We adopt Adam \cite{Kingma2015Adam} as the stochastic optimization method to update the parameters of the framework. When training the upper-level LSTM by Equation \ref{eq:week_likelihood}, we fix the parameters in the lower-level LSTM and only update the parameters in the upper-level LSTM.

\subsection{Experiment Results}

\begin{figure}
    \centering
    \includegraphics[width=0.36\textwidth]{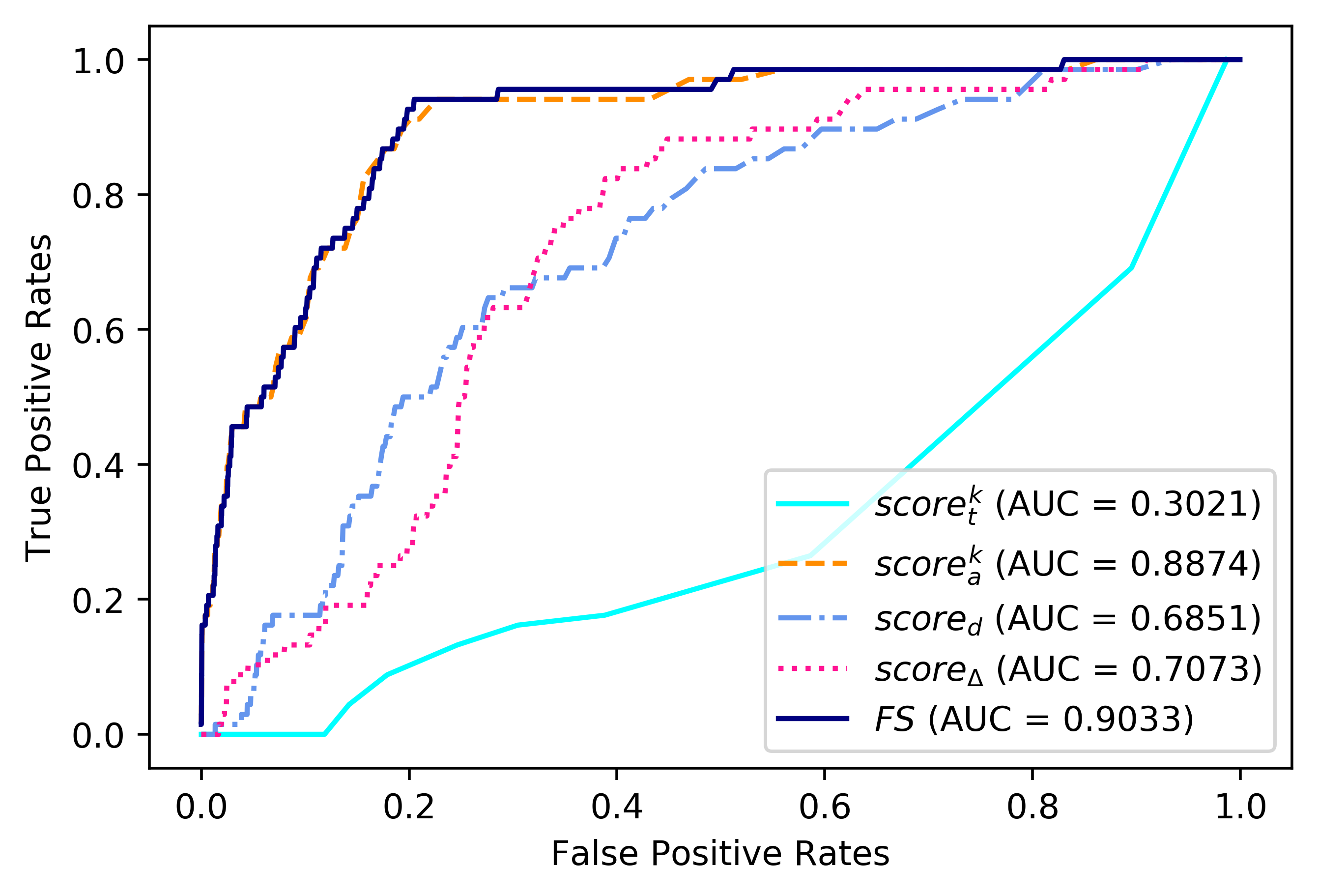}
    \caption{ROC curve of malicious session detection using various fraudulent scores}
    \label{fig:all_sess}
\end{figure}

\begin{figure}
    \centering
    \includegraphics[width=0.36\textwidth]{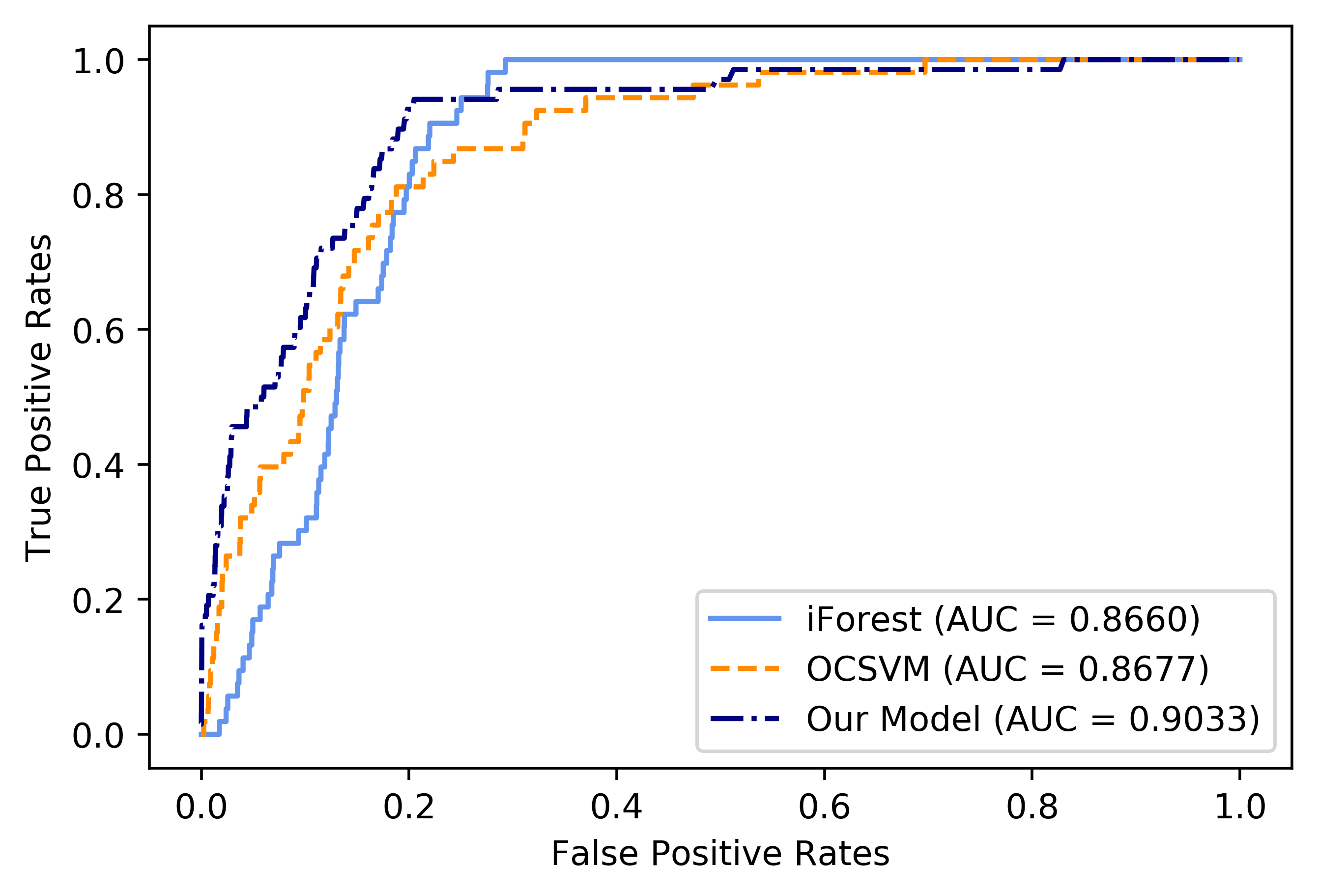}
    \caption{ROC curve of malicious session detection using various approaches}
    \label{fig:baselines}
\end{figure}

     

We aim to detect all the 68 malicious sessions from the totally 142,600 sessions in the testing set. 
Figure \ref{fig:all_sess} shows the receiver operating characteristic (ROC) curves of our model for insider threat detection by leveraging various fraudulent scores. By using each fraudulent score separately, we can notice that the $score_a$ derived from intra-session activity types achieves the highest area under cure (AUC) score, which indicates the activity types of malicious sessions are different from the normal sessions. Meanwhile, the session duration time and inter-session duration time also make positive contributions to the malicious session detection. The $score_{d}$ derived from the session duration time and $score_{\Delta}$ derived from the inter-session duration time achieve good performance with AUC=0.6851 and 0.7073, respectively, which indicates the duration of malicious sessions and inter-sessions are usually different from those of normal sessions. 
We also notice that the $score_t$ based on the inter-activity activity time information does not help much on insider threat detection. The AUC derived from $score_t$ is 0.3021. After examining the data, we find that there is no much difference in terms of inter-activity time information between malicious sessions and normal sessions. Since adopting $score_t$ does not achieve reasonable performance in the CERT Insider Threat dataset, we set $\alpha_2=0$ when deriving the total insider threat detection $FS$.
As a result, our detection model using the total insider threat detection $FS$, which combines all the intra- and inter-session information, achieves the best performance with the AUC=0.9033 when the hyper-parameters in Equation \ref{eq:score} are $\alpha_1=1, \alpha_2=0, \alpha_3=1, \alpha_4=1$. 

Figure \ref{fig:baselines} further shows the ROC curves of our model and two baselines. We can observe that our model achieves better performance than baselines in terms of AUC score. Especially, we can notice that when our model only adopts the activity type information ($score_a$) for malicious session detection, our model is slightly better than baselines in terms of AUC. With further combining the activity time and type information, our model significantly outperforms the baselines with the AUC=0.9033.


\subsection{Vandal Detection}

{\bf \noindent Dataset.}
Due to the limitation of the CERT dataset where the inter-activity duration times are randomly generated, the inter-activity time in the intra-session level does not make contributions to the insider threat detection. 
To further show the advantage of incorporating activity time information, we apply our model for detecting vandals on Wikipedia. Vandals can be considered as insiders in the community of Wikipedia contributors. The study has shown that the behaviors of vandals and benign users are different in terms of edit time, e.g., vandals make faster edits than benign users \cite{kumar2015vews}. Hence, we expect that using the inter-activity time information can boost the performance of vandal detection.

We conduct our evaluation on the UMDWikipedia dataset \cite{kumar2015vews}. This dataset contains information of around 770K edits from Jan 2013 to July 2014 (19 months) with 17105 vandals and 17105 benign users. Each user edits a sequence of Wikipedia pages. We adopt half of the benign users for training and the other half of the benign users and all the vandals for testing. Since user activities on Wikipedia do not have explicit indicators, such as LogON or LogOff, to split the user activity sequence into sessions, we consider user activities in a day as a user session. As a result, the session duration is always 24hrs, and the inter-session duration is 0.  
Therefore, in this experiment, we focus on vandalism session detection with only using information from the intra-session level and adopt the lower-level LSTM shown in Figure \ref{fig:framework} accordingly. Note that we filter out all the sessions with number of activities less than 5. The seq2seq model takes a feature vector as an input and predicts the next edit time and type. In this experiment, we consider the activity type as whether the current edit will be reverted or not. The feature vector of the user’s t-th edit is composed by: (1) whether or not the user edited on a meta-page; (2) whether or not the user consecutively edited the pages less than 1 minute, 3 minutes, or 5 minutes; (3) whether or not the user’s current edit page had been edited before.

\begin{figure}
    \centering
    \includegraphics[width=0.36\textwidth]{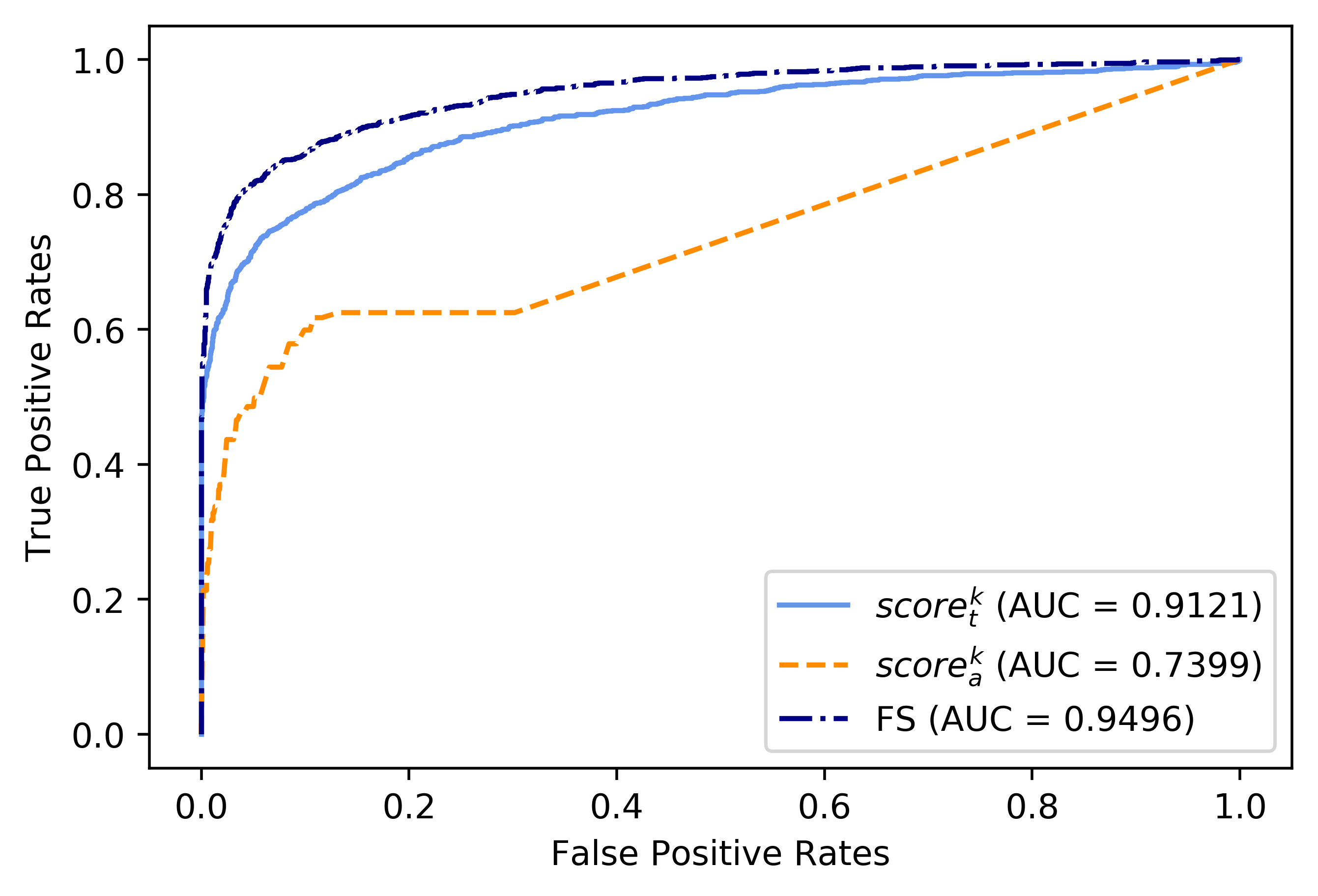}
    \caption{ROC curve of vandalism session detection using various fraudulent scores}
    \label{fig:vews_all}
\end{figure}

\begin{figure}
    \centering
    \includegraphics[width=0.36\textwidth]{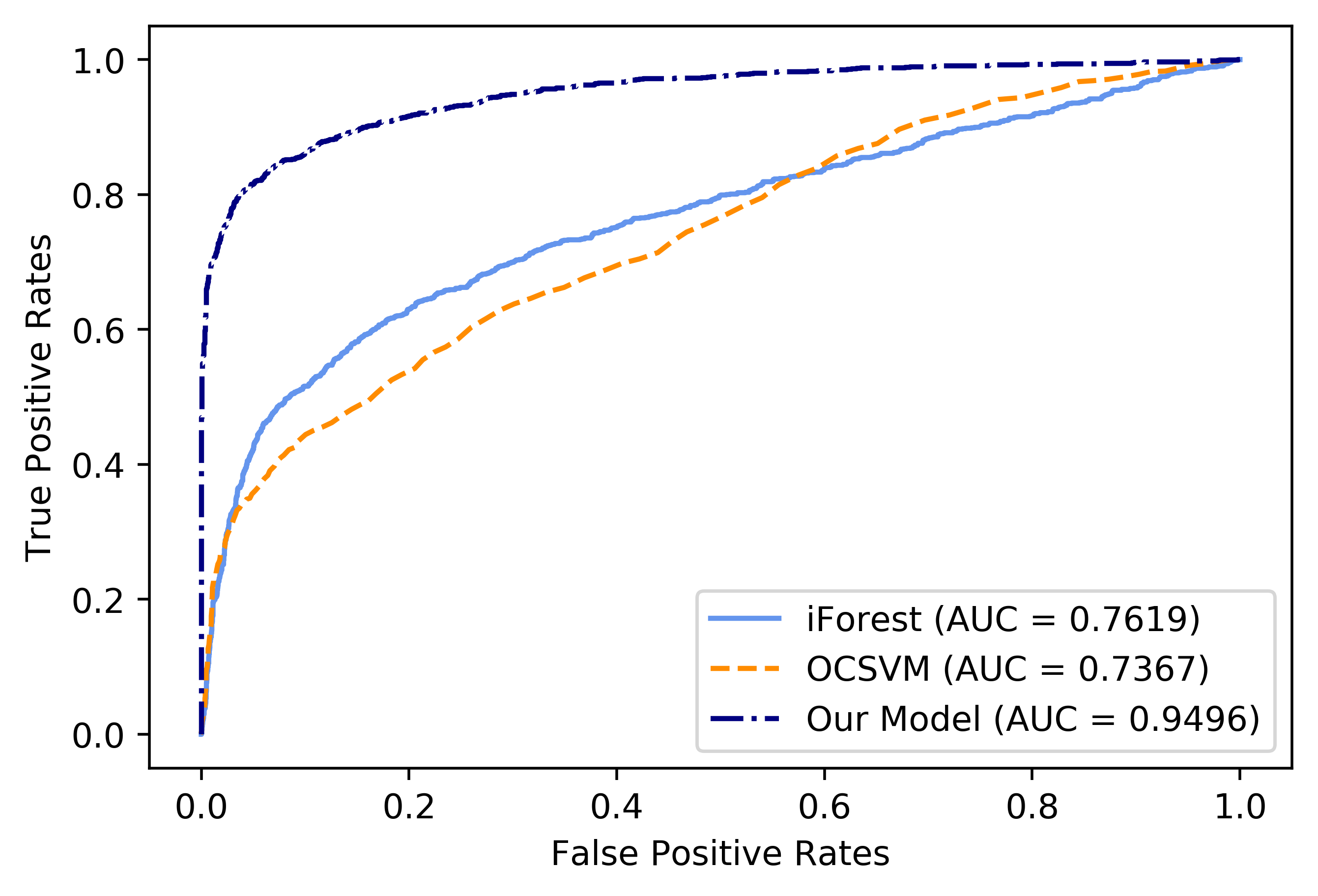}
    \caption{ROC curve of vandalism session detection using various approaches}
    \label{fig:vews_baselines}
\end{figure}


{\bf \noindent Experiment Results.} From Figure \ref{fig:vews_all}, we can observe that only using the inter-activity time information can achieve surprisingly good performance on vandalism session detection with AUC=0.9121, which indicates the inter-activity time information is crucial for vandalism session detection. Meanwhile, adopting the activity type information can also achieve the vandalism session detect with AUC=0.7399. Hence, using inter-activity time information achieves better performance than using the activity type information in terms of AUC. It also means vandals have significantly different patterns in activity time compared with benign users. Finally, with combining the activity type and time information, our model can achieve even better performance with AUC=0.9496. 

We further compare our model with two baselines, i.e., One-class SVM and Isolation Forest. For the baselines, we consider the same features as the seq2seq model and further combine activity types. The value of each feature is the mean value of the corresponding feature in a day. Figure \ref{fig:vews_baselines} indicates that our model significantly outperforms baselines in terms of AUC on the vandalism session detection task. Similar to the results on the CERT dataset, when our model only adopts the activity type information ($score_a$ shown in Figure \ref{fig:vews_all}), the model achieves similar performance as baselines. With considering the activity time information, the performance of our model is improved by a large margin.

\section{Conclusion}
In this paper, we have proposed a two-level neural temporal point process model for insider threat detection. In the lower-level, we combined the seq2seq model with marked temporal point processes to dynamically capture the intra-session information in terms of activity times and types. The upper-level LSTM takes the first and last hidden states from the encoder of the lower-level LSTM as inputs to predict the interval of two sessions and the session duration based on activity history. Experimental results on an insider threat detection dataset and a Wikipedia vandal detection dataset demonstrated the effectiveness of our model. 
\section*{Acknowledgment}
This work was supported in part by NSF 1564250 and the Department of Energy under Award Number DE-OE0000779.

\bibliographystyle{IEEEtran}
\bibliography{Remote}

\end{document}